\documentclass{article}
\usepackage[T1]{fontenc}
\usepackage[utf8]{inputenc}
\usepackage[]{ismir}
\usepackage{amsmath,cite,url}
\usepackage{graphicx}
\usepackage{color}

\usepackage{booktabs}
\usepackage{xcolor}
\usepackage{booktabs}
\usepackage{xcolor}
\usepackage{pifont} % for \ding

\newcommand{\cmark}{\ding{51}} % check
\newcommand{\xmark}{\ding{55}} % cross

\usepackage{array}
\newcolumntype{s}{>{\footnotesize}c}

\usepackage{multicol}
\usepackage{multirow}

\title{D-Composer: Language Modeling for Simultaneous Drum Transcription and Sound Event Decomposition}

\multauthor
   {Patrick O'Reilly \hspace{1cm} Saumya Pailwan \hspace{1cm} Annie Chu}
   {{\bf Jason Smith \hspace{1cm} Bryan Pardo \hspace{1cm}}\\
   Northwestern University, Evanston, USA\\
   {\tt\small patrick.oreilly2024@u.northwestern.edu}
   }

\def\authorname{Patrick O'Reilly, Saumya Pailwan, Annie Chu, Jason Smith, Bryan Pardo}

\def\authorname{P. O'Reilly, S. Pailwan, A. Chu, J. Smith, and B. Pardo}

\usepackage[bookmarks=false,pdfauthor={\authorname},pdfsubject={\pdfsubject},hidelinks]{hyperref}

\begin{document}

\maketitle

\begin{abstract}
Drum beats are often conceptualized and produced as arrangements of repeated single-drum sound events or \textit{one-shots}. Drum Analysis-by-Synthesis (AbS) models combine this structural assumption with reconstruction-based training to perform MIR tasks such as drum transcription, drum source separation, and one-shot extraction, offering a flexible alternative to more traditional task-specific supervision. However, leading AbS approaches struggle to generalize to a broad range of drum timbres (e.g., electronic drum kits), rely on high-quality paired audio-transcription training data, and do not model intra-class timbral variation within a single arrangement (e.g., multiple kick sounds).
To overcome these limitations, we propose \textbf{D-Composer}, an autoregressive language modeling framework that decomposes drum recordings into labeled sound events and their onset times. In drum transcription and one-shot extraction tasks, D-Composer performs comparably to or better than a state-of-the-art AbS system and task-specific baselines while generalizing to a broader range of drum timbres. Our results demonstrate that when equipped with abundant synthetic data and compact acoustic token representations, language modeling provides a simple recipe for unifying transcription and sound event decomposition.

\end{abstract}

\begin{figure}
  \centering
  \includegraphics[alt={D-Composer hero diagram},width=0.9\linewidth]{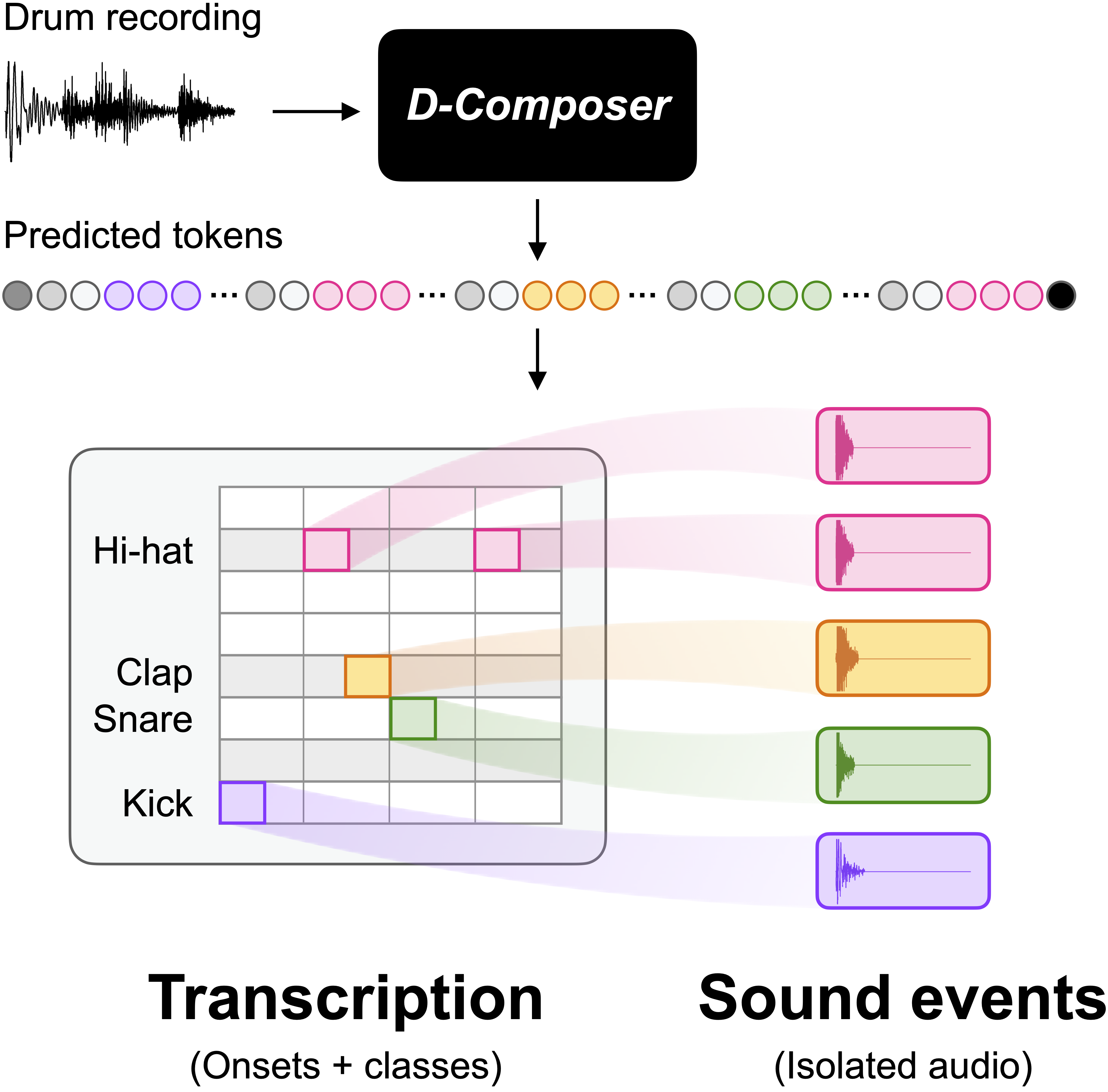}
  \caption{D-Composer maps drum audio to a stream of tokens (see Section \ref{subsec:tokens}) encoding the onset time, class, and isolated audio for all sound events, effectively turning any drum recording into a playable piano roll.}
  \label{fig:hero}
\end{figure}

\section{Introduction}\label{sec:introduction}
Drums are a cornerstone of many musical traditions.  
Within popular Western genres, drums are typically conceptualized (and correspondingly notated and produced) as arrangements of repeated single-drum sound events or \textit{one-shots} spanning drum classes such as kick, snare, and hi-hat \cite{mauch2012corpus, dose, nieto2020audio}. While this discrete structure is important to the creation of music \cite{rhythmicprototypes, lartillot2008multi, rafii2012repeating}, audio recording and mixing processes combine drum sound events into representations (e.g, PCM) that make it extremely difficult to access individual one-shots.
``Factorized" music recording formats that preserve arrangement structure at the note level (e.g., MPEG-4 Structured Audio\cite{mpeg4structured}) have struggled with adoption, in part because there is no reliable way to automatically extract and label note-level audio from recorded tracks.

MIR research has sought to develop methods for recovering this latent structure, including tasks such as \textit{drum transcription} (the mapping of a drum recording to a score-like representation indicating the onset time and class of each sound event), \textit{drum source separation} (the mapping of a drum recording to one recording per drum class, each capturing all sound events of that class), and \textit{drum one-shot extraction} (the mapping of a drum recording to one representative sound event recording per drum class). The resulting task-specific models can recover individual aspects of the latent structure (e.g., event times) but cannot fully decompose a drum recording into a manipulable arrangement of discrete sound events such as a piano roll \cite{adtof, dose}, limiting creative applications.

\begin{figure*}%[hbt!]
  \centering
  \includegraphics[alt={D-Composer system diagram},width=0.9\linewidth]{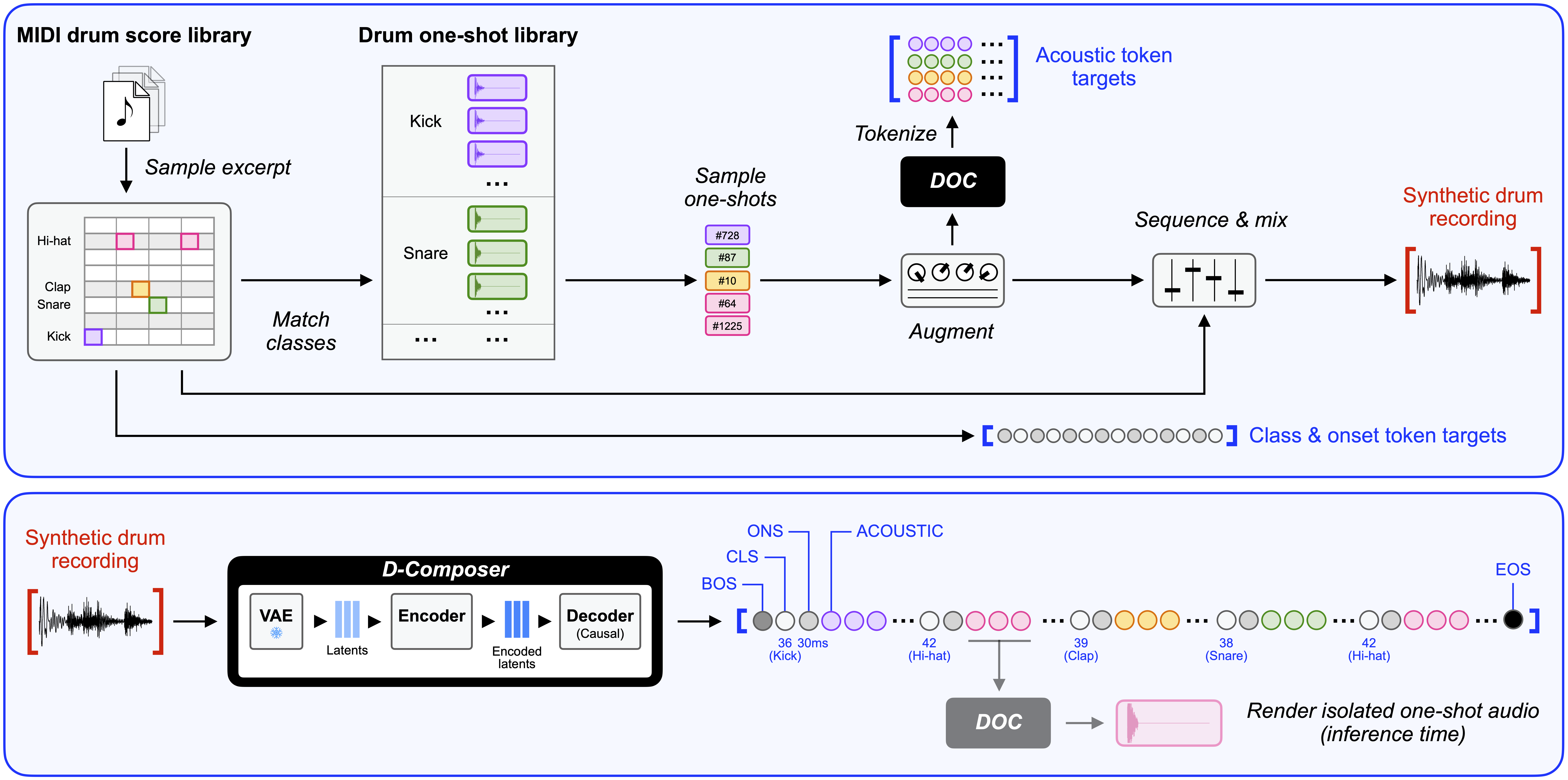}
  \caption{Training of the proposed D-Composer system, including data preparation (top) and target formatting (bottom). Using MIDI drum scores and one-shot data, we construct synthetic drum recordings as model inputs (bracketed in \textcolor{red}{red}) and corresponding drum class, onset, and acoustic tokens for each constituent one-shot as targets (bracketed in \textcolor{blue}{blue}).
   }
  \label{fig:main}
\end{figure*}

Recent work in drum Analysis-by-Synthesis (AbS) has shown that a single model can be trained to perform transcription, separation, and one-shot extraction simultaneously by reconstructing drum audio from a predicted transcript \cite{idm} -- closing the gap towards full decomposition. However, leading AbS approaches struggle to generalize to a broad range of drum timbres (e.g., electronic drum kits), rely on high-quality paired audio-transcription training data, and do not model intra-class timbral variation within a single arrangement (e.g., multiple kick sounds).

To overcome these limitations, we propose \textbf{D-Composer}, an autoregressive language modeling framework that fully decomposes drum recordings into arrangements of discrete sound events. Given a drum recording, D-Composer produces a structured sequence of tokens representing each sound event's onset time (in quantized 10ms intervals), the sound event's drum class (as a MIDI note), and the isolated sound event itself (as a sequence of acoustic tokens that can be mapped to audio through a purpose-built low-bitrate codec). Notably, whereas AbS performs analysis and synthesis in sequence, our framework organizes analysis and synthesis into a \textit{single interleaved process} that produces a combined symbolic-acoustic output.

In drum transcription and one-shot extraction experiments on both real and synthetic recordings, D-Composer performs comparably to or better than a state-of-the-art AbS system and task-specific baselines while generalizing to a broader range of drum timbres, validating the effectiveness of our approach in ``spelling out" drum beats as sequences of sound events. Our contributions are as follows:
\begin{itemize}
    \item A language modeling framework (D-Composer) that fully decomposes a drum beat recording into a sequence of discrete sound events
    \item An ultra-low-bitrate \textbf{d}rum \textbf{o}ne-shot \textbf{c}odec (\textbf{DOC}) that compresses 44.1kHz stereo drum sound events at 0.1kbps, an order of magnitude smaller than  leading high-fidelity codecs \cite{dac, codicodec}
    \item Transcription and one-shot extraction experiments demonstrating that the proposed system's decompositions accurately locate and reconstruct a drum recording's constituent sound events
\end{itemize}

Through this work, we aim to provide a tool for mapping drum beats to piano roll-like representations in which individual sound events can be easily accessed and manipulated to adjust timing and arrangement, extract individual sound sources, or perform other music production tasks. \noindent We provide listening examples and code at \url{https://d-composer.github.io}.

\section{Related Work}

Recent works have applied deep learning methods to drum production and analysis tasks, including drum transcription \cite{adtof, rnntranscription, deepunsup, towardsrealistic, increasingvocab, fewshot, noisetonotes, tapping}, symbolic and audio-domain drum beat generation \cite{learningtogroove, rave, tria}, drum source separation \cite{stemgmd}, drum one-shot extraction \cite{dose}, and drum one-shot synthesis \cite{stylewavegan, gaininversion, drumgan, crash}. Earlier drum transcription methods were tailored to  acoustic drum kit timbres and class taxonomies, limiting their usefulness for timbrally distinct genres such as hip-hop. Recent works have sought to improve the generalization of transcription systems by moving from small annotated datasets of acoustic kit recordings to larger augmented and synthetic datasets \cite{towardsrealistic, increasingvocab, egmd}, and from frame-level discriminative modeling to conditional generative modeling \cite{noisetonotes}. Similarly, recent work in drum source separation and one-shot extraction has used large-scale, high-quality simulated drum data \cite{stemgmd} and generative modeling over synthetic datasets \cite{dose} to achieve strong performance.

As deep learning-based methods for drum analysis and production have improved, Analysis-by-Synthesis (AbS) has emerged as a promising alternative to single-task supervised learning. Initial work in AbS leveraged reconstruction-based training and sparsity regularization to learn \textit{unsupervised} drum transcription \cite{deepunsup}. More recently, Torres et al. proposed the Inverse Drum Machine (IDM), an AbS method that leverages reconstruction-based training and transcript supervision to unify drum transcription, source separation, and one-shot extraction within a single model \cite{idm}. IDM achieves strong performance across these tasks at a size of only 500,000 parameters, demonstrating the benefits of AbS training. However, IDM's design presents three major trade-offs: it predicts dense, sample-level transcripts that require post-processing to sparsify; it yields a single one-shot recording per drum class, preventing the modeling of intra-class timbral variations; and it is trained on simulated drum recordings spanning a narrow range of acoustic drumkit timbres. \textbf{In this work, we propose an alternative framework for training models to simultaneously identify and isolate sound events.} By moving from audio-domain AbS to language modeling over symbolic-acoustic token streams, our framework naturally produces sparse, event-level transcripts; enables modeling of intra-class timbral variations by rendering all sound events individually; and leverages synthetic data to cover diverse acoustic and electronic drum timbres.

Beyond AbS, our proposed method parallels recent works in audio understanding for language models that ground analysis tasks in timestamped transcript-like predictions \cite{listenfirst, tac, whisper}. In particular, our work can be seen as related to the timestamped audio captioning method of Kumar et al. \cite{tac}, which maps an audio recording to a readout of sound event onset times, classes, and descriptive captions. However, our method differs in that descriptive captions are replaced with compact acoustic token sequences that can be used to recover isolated sound events themselves, enabling a range of musical interactions.

The Cerberus method of Manilow et al. also performs simultaneous transcription and separation of musical mixtures, demonstrating the value of joint supervision across multiple tasks \cite{cerberus}. Whereas Cerberus predicts two-dimensional piano roll-like transcript representations in a single step and performs mask-based separation of full instrument classes, we decompose drum recordings into isolated sound events and predict transcripts via language modeling over an event-based token representation. In the transcription space, our ``decomposition'' framing also has precedent in the work of Wu \& Lerch \cite{nmf}, who perform drum transcription via NMF. 

Finally, our deep learning-based mapping of audio recordings to piano roll-like representations has precedent in the Recomposer work of Ellis et al. \cite{recomposer}. However, whereas Recomposer performs transcript-guided resynthesis of general sound mixtures, our approach decomposes a drum beat into isolated sound events and a transcript that arranges them, enabling us to synthesize mixtures through simple summing.

\section{Method}
\label{sec:method}
We next describe the components of the proposed D-Composer system. As illustrated in Figure \ref{fig:main}, D-Composer leverages a simple autoregressive language model (Section \ref{subsec:d-composer}) to map drum recordings to sequences of interleaved transcript tokens and acoustic tokens that respectively locate and characterize each sound event (Section \ref{subsec:tokens}). As ground-truth decompositions are scarce for in-the-wild drum data, we propose to synthesize suitable training data from more widely available drum score and one-shot data sources (Section \ref{subsec:data}). To function efficiently, our system relies on a low-bitrate \underline{d}rum \underline{o}ne-shot \underline{c}odec or DOC (Section \ref{subsec:doc}) to encode isolated sound events as short sequences of acoustic tokens and to correspondingly decode such sequences to audio. The compact representation afforded by DOC allows us to maintain reasonable decomposition sequence lengths even for drum recordings that contain many sound events (as each event requires the prediction of a separate DOC token subsequence). The following sections cover each system component in more detail.

\begin{figure}
  \centering
  \includegraphics[alt={DOC diagram},width=0.99\linewidth]{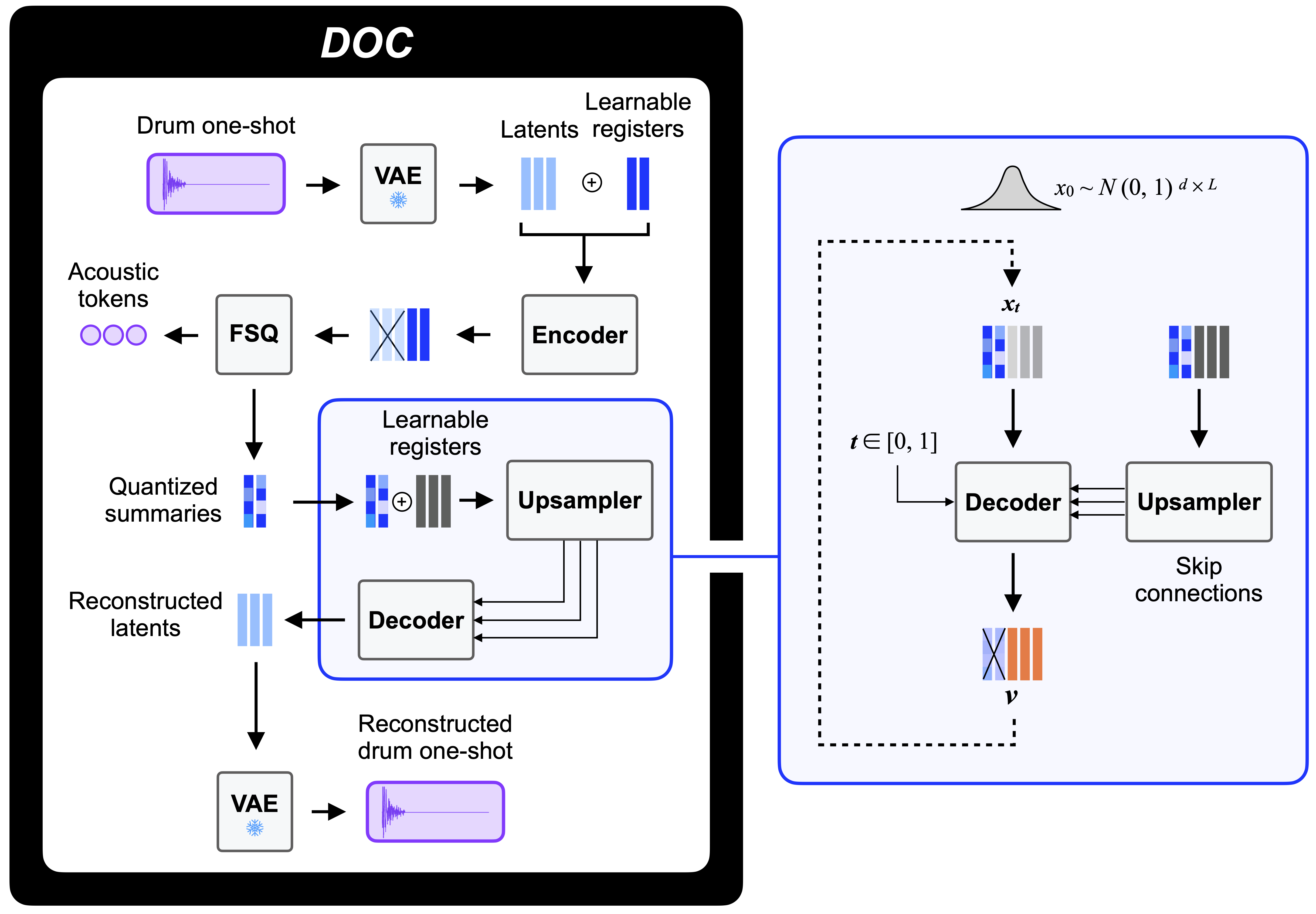}
  \caption{The proposed Drum One-Shot Codec (DOC). FSQ outputs are grouped to obtain the acoustic tokens (shown as purple circles) used by the D-Composer system.}
  \label{fig:doc}
\end{figure}

\subsection{Drum One-Shot Codec (DOC)}
\label{subsec:doc}

Our method predicts sound events as discrete token sequences, leveraging a codec to convert between tokens and drum one-shot audio. Existing high-fidelity codecs \cite{dac, codicodec} require hundreds of tokens to represent a 3-second one-shot, meaning a decomposition of a drum recording containing 50 sound events would require on the order of $10^{4}$ tokens. While feasible for contemporary language modeling approaches, this sequence length imposes significant compute and memory requirements. We therefore train a custom drum one-shot codec (DOC) to map between 3-second one-shot audio recordings and token sequences of length 32, reducing sequence lengths by an order of magnitude and allowing for training on consumer GPUs.

The design of DOC is illustrated in Figure \ref{fig:doc}. Similar to Pasini et al. \cite{codicodec}, we pool an input representation (obtained from a frozen pretrained VAE) to a fixed-length sequence of summary vectors using learnable registers; these summary latents are then quantized per-dimension using Finite Scalar Quantization (FSQ) \cite{fsq} and passed to an upsampling module that restores the input sequence length via additional learnable registers. Following recent works that show the benefit of generative decoding for highly compressed latent representations \cite{sst, spresso, codicodec}, we use a flow-matching \cite{flowmatching} decoder to reconstruct the input latents by refining a corresponding sequence of noise vectors, conditioned on the flow timestep $t$, the summary latents, and the intermediate activations from the upsampler. 

While our approach is compatible with any VAE or invertible representation, we opt for the MelodyFlow VAE \cite{melodyflow} due to its encoding speed and ability to render 48kHz stereo audio. Our encoder, upsampler, and decoder are standard non-causal transformers \cite{transformer} with rotary positional encodings \cite{rope}, 12 layers of dimension 512, and 8 heads per layer for a total of 140m parameters. We use a single summary vector of dimension 128, 5 quantization levels per dimension, and group sets of 4 adjacent dimensions to obtain $\frac{128}{4}$ = $32$ product codes \cite{spectralcodecs} with vocabulary size $5^4 = 625$. DOC is trained end-to-end using only the standard flow-matching loss, achieving roughly full utilization of the FSQ codebook.

\subsection{Interleaved Token Representation}
\label{subsec:tokens}

We predict drum decompositions as interleaved sequences of DOC acoustic tokens (representing the isolated audio of each sound event) and transcript tokens (representing the onset time and class of each sound event). These structured token sequences can be mapped to manipulable representations such as piano rolls by decoding predicted acoustic tokens to obtain one-shots and arranging decoded one-shots according to predicted transcript tokens (see Figure \ref{fig:hero}). 

Following a beginning-of-sequence (BOS) token, each sound event in a decomposition is represented by a drum class (CLS) token, an onset time (ONS) token, and 32 acoustic tokens that can be mapped to audio via DOC (see Section \ref{subsec:doc}); after all sound events, a token sequence is terminated with an end-of-sequence (EOS) token. Sound events are ordered ascending by ONS with a secondary sort on CLS where two events share an onset. ONS tokens are drawn from a vocabulary of 1000 onsets in $[0,  9990]$ms quantized to $10$ms intervals, CLS tokens from a vocabulary of 12 MIDI notes following a simplified general MIDI percussion mapping \cite{generalmidi} (see Table \ref{tab:coarse_vocab}), and acoustic DOC tokens from a vocabulary of size 625. Our token representation is illustrated in Figure \ref{fig:main}. Similar to event-based symbolic music tokenizations \cite{musictransformer, amt}, our representation is naturally sparse (we do not expend tokens on silence) and supports variable numbers of sound events.

\begin{table}[t]
\centering
\scriptsize
\setlength{\tabcolsep}{4pt}
\begin{tabular}{r l r}
\toprule
\textbf{MIDI} & \textbf{Class} & \textbf{Count} \\
\midrule
36 & kick        & 9007 \\
37 & rim         & 220 \\
38 & snare       & 6643 \\
39 & clap        & 1196 \\
42 & hat         & 4331 \\ % 1281 + 130 + 797
43 & tom         & 4191  \\ % 113 + 78 + 163
49 & cymbal      & 9845 \\ % 4128 + 4904 + 23
54 & tambourine  & 350 \\
56 & cowbell     & 129 \\
67 & perc        & 4479 \\ % 4224 + 17 + 150 + 11 + 27 + 40
70 & shaker      & 759 \\
71 & other       & 3708 \\
\midrule
  & \textbf{Total} & 	\textbf{44858} \\
\bottomrule
\end{tabular}
\caption{Drum vocabulary with representative MIDI notes and aggregated one-shot counts.}
\label{tab:coarse_vocab}
\end{table}

\subsection{Training Data}
\label{subsec:data}
Training D-Composer requires drum beat recordings with known ground-truth decompositions. To this end, we synthesize training data by sequencing one-shot recordings according to drum MIDI scores, with the former providing timbral diversity and the latter realistic drum arrangements.

We obtain drum scores from the Lakh MIDI \cite{lmd} and Groove MIDI \cite{learningtogroove} datasets, normalizing notes to match the aforementioned drum vocabularies. We obtain one-shots from the Logic Pro stock library\footnote{\url{https://www.apple.com/logic-pro/}}, the Native Instruments Battery4 library\footnote{\url{https://www.native-instruments.com/en/products/komplete/drums/battery-4/}}, and the Drum and Percussion Kits dataset \cite{organiconeshots}. To match one-shots with scores, we map each one-shot recording to a MIDI note from our drum class vocabulary using available MIDI tags (in the case of some Logic Pro data) or classes inferred from filenames. The resulting one-shot distribution is shown in Table \ref{tab:coarse_vocab}.

The preparation of our training data is illustrated in Figure \ref{fig:main}. To construct a training example, we first sample an excerpt from a MIDI drum score. For all notes present, we sample up to four one-shots in the corresponding drum class to allow for intra-class timbral variation. Using labels present in the Logic Pro and Battery4 data, we require that $40\%$ of training instances sample all one-shots from within the same ``drum kit'' family to ensure exposure to timbrally coherent mixtures. We augment one-shots of each class with random speed change, distortion (hyperbolic tangent and clipping), filtered delay, reverb, and pass-filtering to further boost timbral diversity. Similarly, we apply random time shift, timing jitter, speed change, and velocity jitter to the MIDI excerpt. We then sequence one-shots according to the MIDI excerpt, randomly selecting between sampled variations of each class at every occurrence. The sequenced one-shots are gain-scaled according to the corresponding MIDI velocities and summed, and the resulting mixture is further augmented with reverb, equalization, and dynamic range compression to simulate realistic recording conditions. Target CLS and ONS tokens are extracted from the augmented MIDI sequence, while target acoustic tokens are obtained by passing the selected one-shots through both the class-level and mixture-level audio transformations -- effectively assuming linearity in our audio processing -- and then encoding with DOC. 

\subsection{D-Composer Language Model}
\label{subsec:d-composer}

Our proposed D-Composer system uses a straightforward encoder-decoder transformer language model trained end-to-end with teacher-forced cross-entropy loss. A non-causal transformer encoder maps a drum beat input -- represented, as in DOC, by a sequence of MelodyFlow latents -- to a sequence of continuous vectors that are passed to a causal decoder transformer via cross-attention. The decoder then predicts a token sequence representing a decomposition of the input recording into isolated sound events, their class labels, and their onset times, as detailed in Section \ref{subsec:tokens}. We use the same underlying transformer architecture (positional encoding, etc.) as DOC, with 12 encoder and decoder layers of dimension 512 with 8 heads each for a total of $90$m parameters.

\section{Experiments}

We validate our proposed method's ability to faithfully decompose drum recordings into discrete sound events through a transcription evaluation (measuring the accuracy of predicted drum classes and sound event onsets), a one-shot extraction evaluation (measuring the quality of sound event isolation), and a reconstruction evaluation (measuring the ``round-trip" distance between a sequenced/summed decomposition and its source audio). Additionally, we compare the reconstruction quality and compression factor of DOC against leading audio codecs.

\begin{table}[t]
\centering
\footnotesize
\setlength{\tabcolsep}{3.5pt}
\renewcommand{\arraystretch}{1.05}

\begin{tabular}{lccccc}
\toprule
\textbf{Codec} & \textbf{SR (Hz)} & \textbf{Stereo} & \textbf{Bitrate (kbps)} & \textbf{\#Tok} & \textbf{MSL $\downarrow$} \\
\midrule

DAC \cite{dac}       & 44.1k & \xmark & 7.7 & 2322 & 1.71 \\
CoDiCodec \cite{codicodec} & 44.1k    & \cmark & 2.3 & 512 & 2.25 \\
DOC (ours)       & 48k    & \cmark & 0.1 & 32 & 3.71 \\

\midrule
\addlinespace[0.15em]

{\color{gray} MelodyFlow \cite{melodyflow}}
& {\color{gray} 48k}
& {\color{gray} \cmark }
& {\color{gray} --}
& {\color{gray} --}
& {\color{gray} 1.29} \\

\bottomrule
\end{tabular}

\caption{Codec compression (as measured by \textbf{Bitrate} and encoded token sequence length \textbf{\#Tok}) and reconstruction performance (as measured by multi-scale spectral loss \textbf{MSL}) on $3$-second drum one-shot recordings. DOC operates on continuous latents obtained from the MelodyFlow VAE, which therefore bounds its reconstruction quality.
}
\label{tab:codec_recon}
\end{table}

\begin{table*}[t]
\centering
\small
\setlength{\tabcolsep}{4.5pt}
\renewcommand{\arraystretch}{1.15}

\begin{tabular}{l sss s ssss}
\toprule
\textbf{Model}
& \multicolumn{3}{c}{\textbf{Transcription F1 $\uparrow$}}
& \textbf{One-shot MSL $\downarrow$}
& \multicolumn{4}{c}{\textbf{Reconstruction MSL $\downarrow$}} \\
\cmidrule(lr){2-4}
\cmidrule(lr){5-5}
\cmidrule(lr){6-9}

& E-GMD
& MDB
& Synthetic
& Synthetic
& E-GMD
& MDB
& Synthetic
& FSL10K \\
\midrule

ADTOF \cite{adtof}
& .68/.70/.29/.40/.16
& .95/.84/.10/.68/.29
& .74/.55/.01/.37/.09
& ---
& ---
& ---
& ---
& --- \\

DOSE \cite{dose}
& ---
& ---
& ---
& 5.19/1.37/---/0.68/---
& ---
& ---
& ---
& --- \\

IDM \cite{idm}
& .51/.57/.27/.52/.11
& .84/.72/.10/.43/.10
& .29/.34/.02/.37/.04
& 8.01/3.30/4.37/0.81/1.24
& 10.63
& 17.32
& 22.65
& 27.83 \\

D-Composer
& .55/.59/.20/.42/.13
& .86/.74/.02/.48/.21
& .78/.61/.02/.40/.15
& 5.68/1.26/3.05/0.63/1.17
& 7.87
& 12.27
& 15.83
& 20.39 \\

\bottomrule
\end{tabular}

\caption{Results of our transcription, one-shot extraction, and round-trip reconstruction experiments. Transcription F1 scores and one-shot spectral distances are reported by the ADTOF 5-class vocabulary of KD/SD/TT/HH/CY, while reconstruction spectral distance is computed on full drum recordings. Metrics for tasks or drum classes not supported by a model are indicated with a dash (--).}
\label{tab:evaluation}
\end{table*}

\subsection{Models}

\noindent \textbf{DOC}: We train for $80$k steps at batch size $32$ on a $90\%$ split of one-shots from our combined dataset (see Section \ref{subsec:data}) trimmed/padded to $3$ seconds. We sample flow timesteps from the squared uniform distribution $U[0, 1]^2$, train with EMA decay $0.9999$, and use AdamW \cite{adamw} with base learning rate $10^{-4}$ and exponential decay 0.999996 per step. Total training time on $2 \times$ RTX 4090 GPUs is 14 hours. At inference, we decode acoustic tokens using $5$ linearly-spaced flow steps and Euler sampling.

\noindent \textbf{D-Composer}: We train for $100$k steps at batch size $16$. Following the method discussed in Section \ref{subsec:data}, we train on $4$-second beats synthesized from MIDI sequences of up to 50 events sampled from the Lakh MIDI dataset and Groove MIDI train split, with one-shots from the same training split as DOC. 
We use the same optimization configuration as DOC but omit EMA. Total training time on $2 \times$ RTX 4090 GPUs is 46 hours. At inference, we sample from all D-Composer models with nucleus $0.95$ \cite{nucleus} and constrain predicted token sequences to ensure correct structure (i.e., allow only the sampling of an onset token after a class token, an acoustic token after an onset token, and so on).

\noindent \textbf{Baselines}: On transcription, one-shot extraction, and reconstruction tasks we compare against the state-of-the-art AbS baseline IDM \cite{idm}. We additionally compare against the task-specific ADTOF transcription model \cite{adtof} and DOSE one-shot extraction model \cite{dose}. For our codec reconstruction experiments, we compare DOC against the high-fidelity codecs DAC \cite{dac} and CoDiCodec \cite{codicodec}, as well as the underlying MelodyFlow VAE \cite{melodyflow} (which provides an upper bound on DOC reconstruction quality).

\subsection{Metrics}

We measure transcription accuracy using the standard per-class F1 score with onset tolerance $50$ms \cite{noisetonotes}. To facilitate comparisons, we format transcript predictions using the 5-class drum vocabulary of ADTOF (kick -- BD, snare -- SD, tom -- TT, hi-hat -- HH, cymbal -- CY), onto which the IDM and D-Composer vocabularies can be mapped cleanly.

We measure one-shot extraction quality through the multi-scale spectral loss between predicted and ground-truth one-shots using the same $5$-class mapping, which is supported by both IDM and D-Composer; for DOSE we report the supported subset of BD, SD, HH. Both IDM and DOSE predict a single one-shot per class, while D-Composer predicts a separate one-shot for every sound event in the recording; we therefore select a random representative one-shot prediction of each class from D-Composer's output. We use the MultiScaleSTFTLoss implementation in the \texttt{audiotools} library\footnote{\url{https://github.com/descriptinc/audiotools}} with MSE on magnitudes and spectrogram window lengths of $[2048, 1024, 512, 256, 128, 64]$ as in Hwang et al. \cite{dose}.

Finally, we use the same multi-scale spectral loss to measure reconstruction performance of D-Composer and IDM on full drum beats (i.e. the round-trip distance between input drum recordings and the predicted reconstructions) and to evaluate the reconstruction performance of DOC and baseline codecs on drum one-shot recordings. We note that while our use of spectral losses to quantify model performance follows prior work \cite{dose, idm}, such losses offer only a coarse assessment of audio similarity and do not fully correspond with human perceptual judgements \cite{dpam}. We leave more fine-grained human evaluations to future work and encourage readers to listen to the examples provided on our supplementary webpage\footnote{\url{https://d-composer.github.io}}.

\subsection{Evaluation Data}

We evaluate D-Composer on three tasks across four datasets, totaling $4500$ drum beat recordings. To ensure fair comparisons with DOSE and IDM, we downsample D-Composer outputs from 48kHz to 44.1kHz and downmix from stereo to mono. We normalize all input recordings to $-16$dB RMS, as we find that all evaluated transcription systems exhibit some sensitivity to loudness.

\noindent \textbf{Transcription}: we evaluate transcription accuracy on data from three sources: $1000$ $4$-second excerpts from the Expanded Groove MIDI (E-GMD) dataset \cite{egmd}, which contains aligned transcripts and drum recordings synthesized using acoustic and electronic timbres on a Roland TD-11 drum kit; $1000$ $4$-second excerpts from the MDB-Drums dataset \cite{mdbdrums}, which contains aligned transcripts and real-world drum recordings from the MedleyDB dataset \cite{medleydb}; and $2000$ $4$-second excerpts synthesized in the manner described in Section \ref{subsec:data} from unseen Groove MIDI sequences and one-shots, omitting both MIDI and one-shot augmentations and using a single representative one-shot for each drum class. These data sources therefore cover a wide range of drum timbres and production processes.

\noindent \textbf{One-shot extraction}: We use the same synthetic dataset as for transcription ($2000$ $4$-second excerpts), as the presence of a single timbre per class per recording allows for well-defined ground truth one-shots.

\noindent \textbf{Reconstruction}: We evaluate round-trip reconstruction on the three aforementioned datasets (E-GMD, MDB-Drums, synthetic) in addition to $500$ $4$-second excerpts from the Freesound Loop Dataset \cite{fsl} obtained by filtering for recordings with only ``drum'' instrument tags,  representing a wide range of sonic palettes and audio qualities.

\noindent \textbf{Codec comparisons}: Finally, we compare codec reconstruction quality on $2000$ unseen one-shots from our combined dataset, trimmed/padded to $3$ seconds. As in our D-Composer evaluations, audio is resampled to 44.1kHz and downmixed to mono prior to computing metrics.

We note that our training data does not overlap with ADTOF. We match the train/test split of GMD scores and Logic Pro one-shots used by IDM's StemGMD \cite{stemgmd} in our own data. Our one-shots overlap with the training data of DOSE via the Drum and Percussion Kits dataset \cite{organiconeshots}.

\section{Discussion}

Our experimental results demonstrate D-Composer's ability to accurately locate and isolate sound events within drum recordings, showing the viability of our language modeling approach for drum decomposition.

The results of our transcription evaluation are presented in Table \ref{tab:evaluation}. D-Composer performs comparably to the state-of-the-art AbS IDM model on the heavily acoustic MDB-Drums dataset and partially acoustic E-GMD dataset, and achieves higher average accuracy than both IDM and the task-specific ADTOF system on synthetic data that draws from a broader range of electronic drum timbres. These results do not demonstrate the superiority of D-Composer as a transcription engine -- performance is heavily affected by training domain, and both ADTOF and IDM achieve strong performance at smaller model scales. Rather, these results demonstrate that by training to predict symbolic-acoustic readouts from synthetic data, our framework teaches the model to accurately localize sound events in real drum recordings.

The results of our one-shot extraction evaluation are presented in Table \ref{tab:evaluation}. D-Composer achieves lower average MSL than IDM in every class and the task-specific DOSE model in all but one (KD), indicating its ability to isolate representative one-shot timbres from mixtures. 

The results of our round-trip reconstruction evaluation are presented in Table \ref{tab:evaluation}. D-Composer achieves lower average reconstruction MSL than the AbS IDM model on both in-the-wild, heavily electronic data (FSL10K) and professionally-recorded, majority-acoustic data (MDB), indicating its ability to  produce decompositions that accurately recover the timing and timbre of source recordings.

Finally, the results of our codec reconstruction experiment are presented in Table \ref{tab:codec_recon}. For three-second one-shot recordings, DOC produces token sequences over 10 times shorter than state-of-the-art acoustic codecs at the expense of degraded audio quality, demonstrating a clear compression/fidelity trade-off. While our round-trip reconstruction and one-shot extraction experiments show that this fidelity is sufficient to compete with baseline systems in downstream tasks, the development of a higher-quality low-bitrate codec stands out as a direction for future work.

Taken together, these results demonstrate the benefits of our simple language modeling approach for decomposing sound scenes into constituent isolated sound events and corresponding arrangements. These decompositions can be mapped to piano roll-like representations, allowing for high-level arrangement edits of drum recordings (e.g., removing hi-hats, adjusting speed) and fine-grained adjustments (e.g., timing correction on individual onsets) -- and potentially opening up a range of musical interactions in the process. 

Future work should explore the development of interfaces for musical decomposition systems; improved low-bitrate acoustic codecs; modeling of expanded MIDI drum class vocabularies and additional attributes such as MIDI velocity; and methods for integrating real-world, annotation-free mixtures into the end-to-end training of decomposition systems via e.g. reinforcement learning.

\newpage

\section{Acknowledgements}

This work was supported by NSF Awards Number 2222369 and 2300633. We would also like to thank Mark Cartwright for productive discussions and Fangze Li for assistance in formatting Figures 1-3.

\section{Ethics Statement}

This work proposes a creative tool for decomposing drum recordings into interpretable and manipulable representations, facilitating music production tasks (editing, transcription, one-shot extraction, drum source separation) and encouraging interactive exploration of musical materials. We address ethical implications and limitations of this work below.

The D-Composer and DOC models presented in this paper were trained on transcripts and one-shot recordings from corpora dominated by Western popular music and may perform poorly on out-of-domain timbres and rhythms from other musical traditions. This limitation may be addressed in future work by expanding the training data to reflect a broader set of musical traditions.

Our one-shot training data was sourced from royalty-free commercial libraries (Native Instruments Battery4 and Logic Pro), as existing public-domain one-shot datasets suffer from size and quality limitations. The licenses of our chosen libraries prohibit the redistribution of assets in ``repackaged'' form even in non-commercial contexts. While these licenses do not assert whether AI models trained on assets may constitute such a repackaging, we initially refrain from releasing model weights as we work to resolve this ambiguity. We emphasize that the models trained in this work are intended for educational and research purposes, and we do not propose their use for commercial purposes.

Finally, we acknowledge the environmental impact of the research presented in this work. We estimate the energy consumption of each D-Composer training run (46 hours on $2 \times$ NVIDIA RTX 4090) at $\frac{450 \mathrm{W} \times 2}{1000} \times 46 \mathrm{h} = 41.4 \mathrm{kWh}$, and the energy consumption of each DOC training run (14 hours on $2 \times$ NVIDIA RTX 4090) at $\frac{450 \mathrm{W} \times 2}{1000} \times 14 \mathrm{h} = 12.6 \mathrm{kWh}$ using the reported $450 \mathrm{W}$ power draw of our cards. Model inference was performed on a single GPU for a small fraction of training time, and we therefore omit this cost from our estimates. At roughly 15 DOC runs and 5 D-Composer runs, we estimate the total energy consumption of our experiments at $396 \mathrm{kWh}$, or $175\%$ the mean training cost for ISMIR papers published in 2017--2023 as estimated by Holzapfel et al. \cite{holzapfel2024green}. We note that, while approximate, this energy consumption is likely orders of magnitude below that of leading academic music models \cite{amt, shao}, and is equivalent to about six charges of a Hyundai Ioniq 5 electric car.

\bibliography{ISMIRtemplate}

\end{document}